# ESTIMATION OF TAIL RISK MEASURES IN FINANCE: APPROACHES TO EXTREME VALUE MIXTURE MODELING

by

Yujuan Qiu

A thesis submitted to Johns Hopkins University in conformity with the

requirements for the degree of Master of Science in Engineering

Baltimore, Maryland

May, 2019

# Abstract

This thesis evaluates most of the extreme mixture models and methods that have appended in the literature and implements them in the context of finance and insurance. The paper also reviews and studies extreme value theory, time series, volatility clustering, and risk measurement methods in detail. Comparing the performance of extreme mixture models and methods on different simulated distributions shows that the method based on kernel density estimation does not have an absolute superior or close to the best performance, especially for the estimation of the extreme upper or lower tail of the distribution. Preprocessing time series data using a generalized autoregressive conditional heteroskedasticity model (GARCH) and applying extreme value mixture models on extracted residuals from GARCH can improve the goodness of fit and the estimation of the tail distribution.

**Primary Reader and Advisor**: Dan Naiman

# Acknowledgments

I am especially grateful to my research advisor, Dan Naiman, who encouraged me to conduct mathematical and statistical research, came up with many great ideas and committed to helping me improve my research and writing skills.

# Contents

# List of Tables

# List of Figures

# Chapter 1 Introduction

## 1.1 Motivation

Extreme value theory is used to analyze the likelihood of the extreme deviation from the median of probability distributions and has been implemented to describe the possibility of rare events or abnormal behaviors (Hu 2013). In the finance and insurance industries, the theory is widely considered as the fundamental methodology to estimate tail risk. Statistically speaking, tail risk is a risk measure that the likelihood of an asset or portfolio moves more than three standard deviations from the mean. Tail distribution is the end portion of distribution curves, which corresponds to extremal high or low quantiles estimations of distributions.

The increasing occurrences of rare events in finance like the financial crisis in 2008 have motivated investors to measure and manage tail risk more effectively. Financial data series, however, often exhibit heavy-tailed features with dependence. Using extreme value theory that relies on traditional model assumptions of normality and independence can cause bias and problematic estimation. Many studies have committed to finding out new methods and models for optimizing the estimation of tail distributions. These studies propose to estimate tail distributions by combining non-extremal parts with extremal parts of distributions. Different mixture models and methods have been developed for applications. However, a comprehensive comparative study for these extremal models and methods are still scarce. There is no generalized framework to choose the most appropriate models and methods. The most recent work was conducted by Hu (2011), he compared the performance of different models and analyzed the model flexibility with or without continuous threshold constraints. However, he only focused on model performance without combining proposed methods or approaches for preprocessing data. He did not study in detail for extremal heavy-tailed distributions like the gamma and lognormal distribution, either.

The development of extreme value theory is still at an early stage, but there have been many interesting models and methods proposed by research. An important challenge in the application of extreme value theory is to choose best-fit models and methods. This thesis, therefore, aims to study existing methods and models of the extreme value theory, and compare their performances for different distributions.

## 1.2 Thesis Objectives

The main objectives of this thesis are to study and implement existing extreme value mixture models and methods into simulated distributions and financial data series and to investigate whether there is a generalized framework to choose best-fit models or methods for different distributions or datasets.

With the increasing high volatility of current financial markets, an accurate estimation of extremal and unusual events becomes more critical to manage risk and minimize the loss. Financial market data is simulated as a process of a random walk with uncertainties and volatilities. An efficient model should capture those uncertainties and perform analysis with all available information. Traditional extreme models include parametric, semi-parametric, and non-parametric approaches, but they only focus on the distribution above the threshold and ignore the distribution below that threshold.

Current existing extremal mixture methods have made many contributions to create more flexible models to measure tail risk and include all available data information. The method to preprocess data before applying different models is not consistently developed either. Each extremal mixture method has its advantages and disadvantages, but it's also uncertain which methods would perform better for datasets with different features.

Extremal mixture models combine non-extremal and extremal portions of a distribution. The parametric form of mixture models uses a threshold to separate extreme and nonextreme domains. While making the selection of the threshold more flexible, the question arises whether this method can provide enough flexibility if the non-extremal portions are mis-specified. Considering the potential drawbacks of existing extremal mixture models, this thesis aims to analyze each model and method's advantages and disadvantages, and test which fits different dataset better. The primary purposes are summarized below:

1. Derive a generalized framework for choosing a best-fit model and method for different datasets

2. Study and compare the advantages and disadvantages of current existing extremal mixture models and methods

3. Summarize the pros and cons of extremal mixture models and methods and propose potential suggestions for improvement.

## 1.3 Thesis Structure

Chapter 2 reviews and summarizes the key concepts and materials on methods of risk measurements, time series models, threshold diagnostics methods, and the extreme value theory.

Chapter 3 extends Chapter 2's statistical and mathematical concepts into different extreme value mixture models and methods, giving an overview of existing models and methods to improve the estimation of tail distribution using extreme value mixture models and methods.

Chapter 4 introduces the simulated and real-life data. Chapter 5 provides a very detailed simulation study that compares the performance of different models and methods and summarizes the findings. Chapter 6 shows the application results on 4 financial and insurance datasets. Chapter 7 summarizes the results of this thesis and provides a conclusion.

# Chapter 2 Background

## 2.1 Return Series

Let $P_t$ and $P_{t-1}$ denote the daily closing prices of an asset $i$ at time $t$ and at previous trading period, $(t-1)$. The arithmetic return on the asset $R_i$ then is given by

$$R_t = \frac{P_t - P_{t-1}}{P_{t-1}}.$$

However, arithmetic returns do not add to multi-period returns. To continuously compound and add multi-period returns, the logarithmic return

$$R_t = \ln \frac{P_t}{P_{t-1}}$$

is often used. The loss return can be expressed as

$$X_t = -R_t.$$

## 2.2 Risk Measurement

Value at Risk (VaR) and Expected Shortfall (ES) are two common risk measurements of an investment portfolio. VaR measures the maximal loss for investment during a period given a quantile of probability. While VaR provides a single number indicating the maximal loss for a given probability, it does not give any information regarding the tail portion of the distribution. Expected shortfall addresses this shortcoming of VaR, and it quantifies the expected losses that occur beyond the VaR threshold. ES measures the average of all losses that are greater or equal than VaR.

## 2.2.a Value at Risk

VaR measures the maximal loss of a portfolio over a defined period for a given confidence interval. For example, if the 99% VaR on a portfolio is $1000 million for a given week, there is only 1% chance that the value of the portfolio will drop more than $1000 million within a week. Mathematically speaking, the VaR of X given a confidence level $\alpha$ is the smallest number $x'$ such that the probability that $X \geq x'$ does not exceed $(1 - \alpha)$ (Artzer et al. 1999)

$$VaR_\alpha(X) = \inf\{x \in R : P(X > x) \leq 1 - \alpha\} = \inf\{x \in R : F_X(x) > \alpha\}.$$

If $F_X$ is continuous, $VaR_\alpha(X)$ is $F_X^{-1}(\alpha)$, representing zth quantile of the distribution of X. VaR can be estimated from the order statistics of historical asset loss values. Suppose that $X_{(1)}, X_{(2)}, X_{(3)}, \ldots, X_{(n)}$ are order statistics for the sample $X_1, X_2, X_3, \ldots, X_n$ of loss values such that $X_1, X_2, X_3, \ldots, X_n$ are identically independently distributed. If $X_{(1)} < X_{(2)} < X_{(3)} \ldots < X_{(n)}$, then the VaR at $\alpha$ level can be estimated

$$VaR_\alpha(X) = X_n(\alpha).$$

For the continuous case, non-parametric methods are commonly used to estimate VaR. Historical simulation (HS) assumes a similar future and past risks. It "simulates" the cumulative distribution function (CDF) of asset returns over past periods without any particular distribution assumption. By weighing returns of assets over the periods equally, HS estimates the current VaR values based on all past periods. Nevertheless, HS has two drawbacks. With the assumption that future and past risks are identical, HS fails to capture the rapid market changes. Leading factors that affect market activities such as investors' confidence, the macroeconomic environment, and regulation policies change over time. Their corresponding risks are not identical. Moreover, data that is further from the present should have a diminishing effect on the present. Equal-weighted returns over time violate this diminishing effect rule.

The Monte Carlo simulation technique builds on the stochastic feature of financial datasets. By using stochastic simulation, Monte Carlo generates a set of possible future prices based on chosen models. These models, however, are typically based on a particular distribution assumption for the data. For example, a model might assume data to be normally distributed. Real data might rarely be perfectly normal distributed or precisely described by any particular distribution. Using Monte Carlo simulation can still underestimate or overestimates the risk.

### 2.2.b Expected Shortfall

Expected shortfall (ES) helps measure the performance of tail distribution. ES is the average loss of a portfolio that exceeds the threshold of VaR given a certain probability, that is, $E[X|X > VaR_\alpha(X)]$ (Rydell 2013)

$$ES_\alpha = \frac{1}{1-\alpha} \int_{\gamma=\alpha}^{1} VaR_\gamma X \mathrm{d}(\gamma)$$

## 2.3 Time Series Models

Time series models are constructed to extract statistical and analytical information from time series data. Different models show diverse stochastic characteristics of the series. A broad class of models can be built using three key building bricks: autoregressive (AR) components, moving average (MA) components, and integration (I) components, leading to what are referred to ARIMA (models). Among non-linear time series model, the autoregressive conditional heteroscedastic (ARCH) model was proposed to analyze data series with time-varying volatility clustering.

## 2.3.a Autoregressive Model (AR)

An autoregressive model (AR) represents a stochastic process to describe the time-varying dependence between the variable of interest and its past values (Hellman 2015). Because of the serial autocorrelation between time series data, an AR model can model such dependence and forecast the conditional mean. That is to say, the AR model describes how $i$ past observations influence the current conditional mean. The specific mathematical form of the AR model is

$$X_t = \phi_0 + \sum_{r=1}^{r} \phi_i X_{t-i} + \epsilon_t.$$

where $\phi_i$ is a constant, and $\phi_0, \phi_1,…,\phi_i$ are the coefficients of the model and $\epsilon$ is a white noise process such that $\epsilon \sim iid\ N(0, \sigma_t^2)$.

## 2.3.b ARCH and GARCH models

In 1982, Engle proposed using the autoregressive conditional heteroscedastic (ARCH) models to model time series that shows time-varying volatility clustering. AR models describes concerns the conditional mean structure of time series data assuming constancy of variance across the time. However, in practice, the conditional variance may vary from the current and past values of process, and, as such, the conditional variance is itself a stochastic process, often referred to as the conditional variance process (Cryer & Chan 2008). For instance, the conditional variance of a stock is often higher following a period of violent price movement period than a period of stable price movement. This pattern of volatility is called volatility clustering. In general, the conditional variance of the return of an asset in finance refers to a risk measure such as VaR.

ARCH is used to model the conditional variance of time series, particularly for time series that has short periods of increased variation. We follow the notations from the book wrote by Cryer & Chan 2008. Let time series $r_t$ represents the daily asset returns, ARCH (1) suggests that the return series $r_t$ split into a stochastic piece $\epsilon_t$ and a conditional standard deviation $\sigma_{t|t-1}$

$$r_t = \sigma_{t|t-1}\epsilon_t.$$

The conditional variance $\sigma_{t|t-1}$ with ARCH (1) is

$$Var(r_t|r_{t-1}) = \sigma_{t|t-1} = \alpha_0 + \alpha_1 r_{t-1}^2$$

where $\alpha_0 \geq 0$ and $\alpha_1 \geq 0$. To extend ARCH (1) in the general case, the conditional variance is modeled as a linear function of past residuals, $\epsilon^2$, which is named as ARCH-terms. The general ARCH(p) is defined as

$$\sigma_{t|t-1}^2 = \alpha_0 + \sum_{i=1}^{p} \alpha_i r_{t-i}^2.$$

ARCH model forecasts the future conditional variance only involves the most recent returns. In practice, the accuracy of forecasting improves by including all past variances.

Generalized autoregressive conditional heteroscedastic (GARCH) model was introduced by Bollerslev in 1986 to add GARCH terms. The general GARCH model takes the form to be

$$\sigma^2_{t|t-1} = \alpha_0 + \sum_{i=1}^{p} \alpha_i r^2_{t-i} + \sum_{j=1}^{q} \beta_j \sigma^2_{t-j|t-j-1},$$

where $\alpha_0 \geq 0$, $\alpha_1 \geq 0$, and $\beta_j \geq 0$ is the coeffiect of GARCH.

## 2.4 Volatility Clustering and Persistence

Volatility clustering represents a phenomenon that small and large changes in a series tend to occur in clusters. To be more specific, a high violent period of movement is likely to follow with a large change, and a smooth period of asset movement is followed by the smaller volatility. Typically, volatility clustering indicates that time series are not normally distributed and suffer from dependence. Normality and dependence are crucial assumptions for traditional extreme value and time series model. If these assumptions break, the accuracy and forecasting power of these traditional models are weakened.

The following figure shows an example of GARCH (1,1) process that shows a volatility clustering. During the periods around t=200 and t=380, the change and fluctuation of the asset (y-axis) are both smaller than the periods around t=180 and t=650. It indicates that the bigger change of the asset follows with the higher volatility.

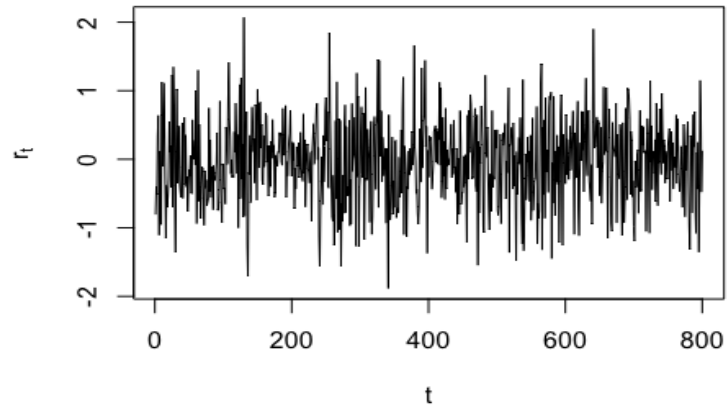

Figure 2.1: GARCH(1,1) Example

## 2.5 Leverage Effect in Financial Time Series

The leverage effect refers to the generally negative correlation between an asset's returns and its changes of volatility. Specifically, a downward movement of an asset is followed by higher volatility and an upward movement of an asset tends to follow by lower volatility. The term leverage refers to an economic interpretation developed by Black (1976) and Christie (1982): as asset prices decline, companies become more leveraged because the relative value of their asset of debt rises over that of their equity. As a result, companies expose higher default risk thus becoming more volatile.

The risk of an asset can be quantified by the expected shortfall or VaR to estimate the tail distribution of the financial series. The leverage effect shows the varying pattern of the volatility, indicating non-constant volatility over periods. Again, a new framework that adjusts for the varying volatility needs to develop to increase the accuracy of risk measurement and reduce the biases.

## 2.6 Extreme Value Theory

Extreme value theory is used to analyze the likelihood of the extreme deviation from the median of probability distributions and has been implemented to describe the probabilities of rare events or abnormal behaviors (Hu 2011). The key idea behind the extreme value theory is to extrapolate the information from the probabilities of extremal portions of distributions rather than centers of distributions.

Generally speaking, there are three approaches to apply EVT: parametric, semi-parametric, and non-parametric methods. The main parametric methods are block maxima and peaks over threshold (POT). The block maxima approach is based on the Fisher-Tippett-Gnedenko theorem. c (Ferreira 2015). In the peaks over threshold approach uses the Pickands-Balkema-de Haan theorem that addresses that observations exceeding a certain high threshold is approximately a Generalized Pareto distribution (GPD) (Pickands 1971). The estimation of the likelihood of rare events then concentrates on the conditional distribution above the threshold by implementing Generalized Pareto distribution. An detailed review of Generalized Pareto distribution is provided in the section 2.6.b.

### 2.6.a Generalized Extreme Value Distribution (GEV)

Let $X_1, X_2, \ldots X_n$ be a series of identically, independently distributed random variables with common distribution $F_X(x)$. Then the distribution of maximum is:

$$M_n = Max(X_1, X_2, \ldots, X_n) = P(M_n \leq x) = P(X_1 \leq x, X_2 \leq x, \ldots, X \leq x) = F^n(x)$$

In the block maxima approach, the Fisher-Tippett-Gnedenko theorem states that the maximum distribution converges in distribution to a non-degenerated limiting distribution which belongs to either Fréchet, Gumbel or Weibull family.

With standardized variable $s = \frac{x-\mu}{\sigma}$ and $\xi$ that describes the fatness of the tail,

**The Frechet family:** $\xi > 0$

$$F(x) = \begin{cases} \exp\left[-\left(1 + \xi s^{\left(-\frac{1}{\xi}\right)}\right)\right] & if\ x > \frac{-1}{\xi} \\ 0 & Otherwise \end{cases}$$

**The Gumbel family:** $\xi = 0$

$$F(x) = \exp[-\exp(-s)] \quad -\infty < x < \infty.$$

**The Weibull family:** $\xi < 0$

$$F(x) = \begin{cases} \exp\left[-\left(1 + \xi s^{\left(\frac{-1}{\xi}\right)}\right)\right] & if\ x < \frac{-1}{\xi} \\ 0 & Otherwise \end{cases}.$$

The parameters $\mu$, $\sigma$, and $\xi$ are estimated using the maximum-likelihood distribution. Using the block maxima method, we divided time series into equally sized blocks where the maxima for each block is determined by the framework above. To describe three families under a general umbrella, we can use the General Extreme Value Distribution (GEV) that takes the form:

$$G_{\mu,\sigma,\xi} = \exp[-(1 + \xi s^{-1/\xi}]$$

where $\xi \neq 0$ and $s = \frac{x-\mu}{\sigma}$.

## 2.6.b Generalized Pareto Distribution (GPD)

In the Peaks over threshold (POT) method, we estimate the conditional distribution that is beyond a certain threshold u of the original distribution. Unlike the block maxima method, the Peaks over threshold method addresses that the exceedances of the distribution belong to the Generalized Pareto distribution (GPD), $G(x|u, \sigma_u, \xi)$. More importantly, the POT method is more efficient when all extreme data are available since it makes use all of them.

Let $X_1, X_2, \ldots X_n$ be a series of identically, independently distributed random variables with common distribution $F_X(x)$. $u$ is the threshold level and $X_k - u$ represent the exceedance level. Then the conditional distribution of exceedance over u threshold is:

$$F_u(x) = P(X - u \leq x | X > u) = \frac{F(x+u) - F(u)}{1 - F(u)}.$$

Here $F_u(x)$, represents the conditional probability that a loss exceeds u by no more x given the threshold at u. Based on Pickands-Balkema-de Haan theorem, $F_u(x) \approx G(x|u, \sigma_u, \xi)$, where $u \to \infty$. The distribution of GPD, $G(x|u, \sigma_u, \xi)$, are:

$$G(x|u, \sigma_u, \xi) = \begin{cases} 1 - \left[1 + \xi\left(\frac{x-u}{\sigma_u}\right)\right]^{-\frac{1}{\xi}} & if\ \xi \neq 0 \\ 1 - \exp\left[-\frac{x-u}{\sigma_u}\right] & if\ \xi = 0 \end{cases}$$

and where $x > u$, $\sigma_u > 0$, and $[1 + \xi(\frac{x-u}{\sigma_u})] > 0$.

## 2.7 Extreme Value Mixture Model

Extreme value theory has been a fundamental tool to measure and predict tail distributions. Traditional Generalized Pareto distribution discards non-extreme data and perform statistical analysis on extremal data to fit Generalized Pareto distribution. Ignoring non-extremal data has two advantages (Hu 2013):

1. Ignoring the non-extremal data that is under the threshold reduces the noise in the estimation process of the likelihood of rare events.

2. If unusual and extremal events are of interest, there is little reason to consider non extreme data since non-extremal and extremal are produced by different processes.

Ignoring non-extremal data, however, has two concerns. There is no general framework to choose thresholds. If the threshold is mis-specified, the estimation of tail distribution is biased and inaccurate. The extremal data is also very scare, and the limited sample size makes it hard to conduct precise measurements.

For the last decades, there has been increasing interests in extreme value mixed models that provide an innovative approach for the estimation of the threshold and use all the data for parameter estimations (Hu 2013). As shown in the Figure 2.2, this method includes all data in the distribution and divides non-extremal and extremal portions of distributions into bulk and tail models. Bulk models describe the behavior of non-extremal parts of distributions and tail model estimates the extremal parts of distributions. These two models simultaneously capture the distribution under and over the threshold, so they include all available information in the process of estimation.

**Threshold selection**: Traditional Generalized Pareto distribution fitting methods require one to choose a fixed threshold to split bulk and tail models. The selection of a threshold is a tradeoff between bias and variance, according to Coles (2001). If the threshold is too low, the asymptotic features underlying the derivation of the GPD model are violated; If the threshold is too high, there would be too few exceedances to measure and would lead to a high variance estimation (Hu 2011). Moreover, there might be multiple thresholds that fit same datasets, thus creating an inconsistent tail distribution and return estimations. Unlike traditional parametric or non-parametric model, extremal mixture model methods treat the threshold as a parameter to find the best fit estimation. Therefore, the uncertainty and inconsistency of the threshold selection is improved by making the inference of threshold parameter.

**Use of all information** Both traditional Generalized Pareto distribution and block maxima methods ignore the distribution under threshold. While these methods have advantages, the extremal mixture model method includes all information to find a flexible bulk model and tail model that fit both non-extremal and extremal data simultaneously. Although non-extremal data does not contribute much to the tail distribution analysis, it still includes important information to acquire a comprehensive overview for the whole dataset. Extremal mixture model therefore allows to discover the best fit "mixed model" to include all available information to dataset.

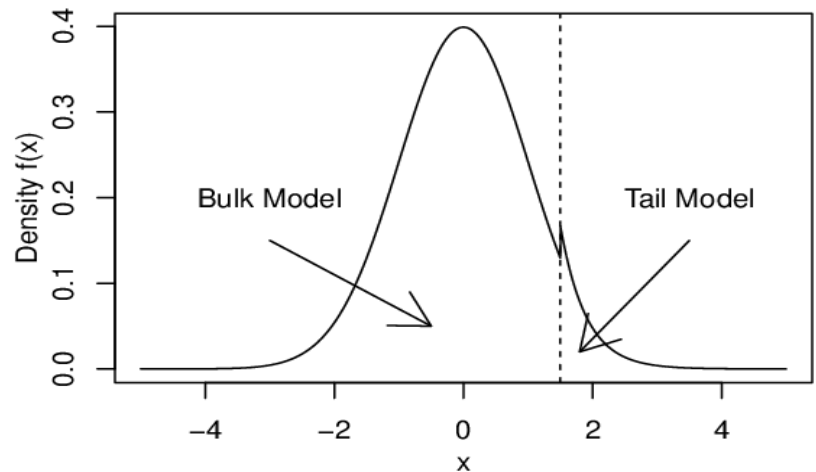

Figure 2.2: Extreme Value Mixture Model (Source: Research Gate)

## 2.8 Threshold Choice - Graphical Diagnostics

Traditionally, thresholds are chosen before fitting the observations, an approach that does this is referred as the fixed threshold method. Graphical diagnostics are widely used to identify appropriate thresholds, balancing the bias and variance of the model fit. Coles (2001) summarizes two conventional approaches to conduct fixed threshold graphical diagnostics:

- mean residual life (or mean excess) plots,

- threshold stability plots.

The advantages of using graphical diagnostics are that they allow researchers to investigate and estimate the data graphically, comprehend a dataset's features and patterns, and analyze goodness of fit. However, this traditional approach suffers two drawbacks. First, choosing a threshold requires expertise and the selection can be very subjective. Due to this subjectivity, research results may be hard to replicate for further validation and study. Secondly, graphical diagnostics can be time-consuming for a large number of datasets. It is common for researchers to analyze many datasets like return levels for different investment portfolios. Although these thresholds are all consistent with the general guidelines for threshold choice using graphic diagnosis, they provide very different tail extrapolations (MacDonald 2011).

Over last decades, many proposals have appeared in the literature for providing more flexible frameworks for the estimation of extreme value distribution. They mostly use asymptotic optimality-based arguments under various population distribution assumptions. However, whether those innovations, flexible methods are superior to the fixed threshold method is unclear. In the simulation and application sections, we conducted a comparative study regarding traditional and extreme value models.

### 2.8.a Mean Residual Plot

Mean residual plots have been introduced by Davison & Smith in 1990 to make a threshold choice by using the expectation of the GPD excess. Let X be a random variable with distribution function F and right endpoint $x_F$. For a fixed $u < x_F$, define

$$F_u(x) = P(X - u \leq x | X > u)$$

where $x \geq 0$, i.e the excess distribution function of random variable X over the threshold u. Then for any higher $v > u$ and $\xi < 1$ the expectation of the GPD excess is

$$E(X - u \mid X > u) = \frac{\sigma_u + \xi v}{1 - \xi}.$$

Note this is a linear function of v. The excess over the threshold is critical to the analysis of extreme events, and this fundamental idea is widely used in many fields. For example, $F_u$ is known as the excess-life or residual life distribution function in reliability theory and medical statistics. In an insurance context, $F_u$ is usually referred to as the excess-of-loss df (Embrechts and Mikosch 1997).

For example, Figure 2.3 gives Mean Excess life graph of Fort Collins total daily precipitation from *extRemes* R package. Three potential thresholds are suggested using maximum likelihood parameter, 0.395, 0.85, 1.2 all could be the potential thresholds following the general guideline.

## 2.8.b Threshold Stability Plot

Threshold stability plots are rather straightforward. It plots the maximum likelihood function of the Generalized Pareto distribution scale and shape parameter $\xi$ against a set of possible thresholds. If the Generalized Pareto distribution is an appropriate model for a threshold u, all v>u will also be suitable. The shape and modified scale then become constant. According to Coles(2001), the modified scale parameter is $\sigma_u - u\xi$. "If the exceedance X above threshold u follow $GPD(\sigma_u, \xi)$, then for all X such that $X > v$ where v > u also follows $GPD(\sigma_u, \xi)$. The shape parameter of GPD is threshold invariant, but the scale paramter is a linear function of the threshold differences", based on Hu (2018). By detecting the stability of shape parameter, researchers can use threshold stability plot to choose a fixed threshold.

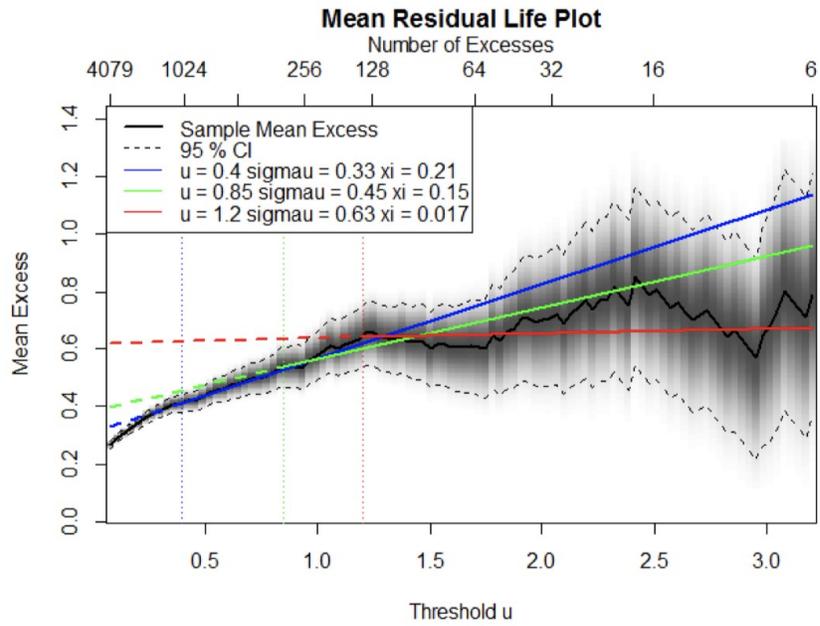

Figure 2.3: Mean Residual Life Plot Example (Hu 2013)

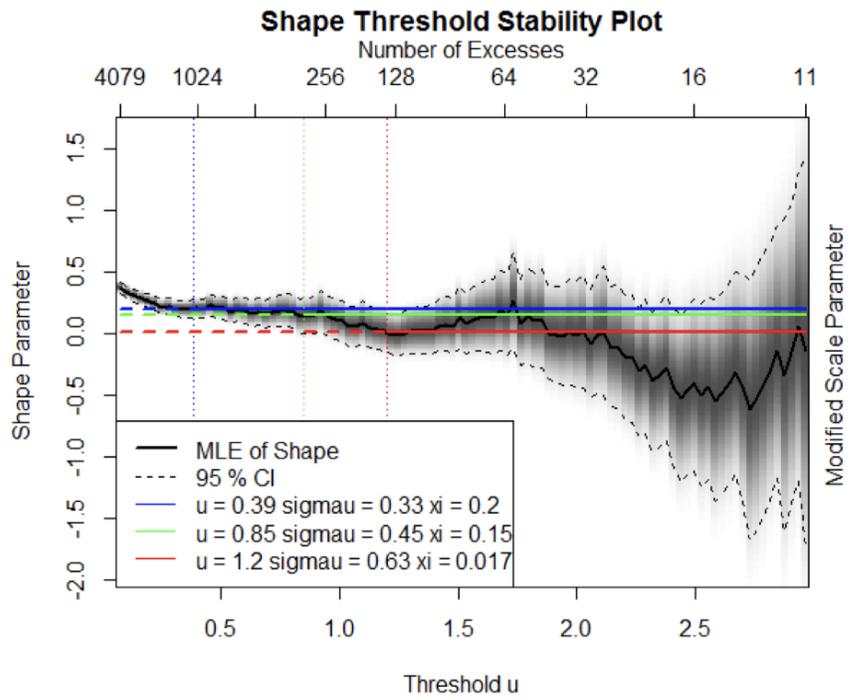

Figure 2.4: Shape Threshold Stability Plot Example (Hu 2013)

# Chapter 3 Extreme Value Mixture Models

In this chapter, key extreme value models and methods are discussed. This chapter compares the advantages and disadvantages for different models to deepen the understanding of the tradeoff among different models and methods' usage.

To analyze tail distributions, the most well-known preprocessing method is proposed by McNeil and Frey (2000). They propose to use a combination of GARCH and extreme value models. Traditional GARCH models capture the volatility clustering and the persistence but fail to account for the fatness of tails. Extreme value methods related to GPD modeling that focuses on the distribution above or below the threshold is superior to estimating the tail risk, but fails to capture volatility clustering. To integrate the advantages of two models, recent studies have developed mixture models to combine GARCH and extreme value theory (EVT) methods and models together with an optimized threshold choice.

## 3.1 Extreme Value Mixture Models

### 3.1.a Extreme Value Mixture Model Overview

In this section, we will review key extreme value mixture models in details and their underlying advantages and disadvantages.

The bulk model has parametric, semi-parametric, and non-parametric forms. The main parametric models that are currently available in the literature are:

- Normal with GPD for upper tail (Cabras and Castellanos 2011)

- Gamma with GPD for upper tail (Behrens et al. 2004)

- Weibull with GPD for upper tail (Behrens et al. 2004)

- Log-normal with GPD for upper tail (Solari and Losada 2004)

- Beta with GPD for upper tail (MacDonald 2012)

- Normal with GPD for both upper and lower tails (Zhao, Scarrott, Reale, and Oxley 2010; Mendes and Lopes 2004)

**Semi parametric models include:**

- The finite mixture of gammas with GPD for the upper tail (do Nascimento, Gamerman, and Lopes 2012)

Non-parametric models with kernel density methods are:

- Standard KDE using a constant bandwidth with GPD for the upper tail (MacDonald et al.2011)

Other models:

- Hybrid Pareto (Carreau and Bengio 2009)

### 3.1.b Parametric Forms of Mixture Model

Behrens (2004) developed a parametric model that takes threshold as a model parameter to fit observations below it and use GPD modeling to fit observations above it. This model uses all observations to estimate model parameters, including the threshold parameter.
Bayesian inference is used to estimate the prior and posterior distributions. By comparing
Behrens' approach with classical models, the poor performance of the scale parameter
estimates is attributed to increasing uncertainty in the Bayesian inferences.

Behrens's approach distribution of the model is defined as

$$F(x|u, \sigma u, \xi, \eta) = \begin{cases} H(x|\eta) & if \quad x \leq u \\ H(u|\eta) + [1 - H(x|\eta)]G(x|u, \sigma u, \xi) & if \quad x > u \end{cases} \quad (1)$$

Where $H(x|\eta)$ is the conditional bulk distributions of X under the threshold taken to be
gamma, Weibull, or normal. Then u is the threshold parameter, $\eta$ parameterizes the marginal
distribution under u, and $\sigma_u$ and $\xi$ are scale and shape parameter from the GPD
distribution. Behrens' model is very straightforward, and it benefits from using Bayesian
inference to compensate limite ford extreme data from expect prior information. However,
Behrens' model have some drawbacks due to the discontinuity or separation point between
the bulk and tail models. The discontinuity may ignore significant information from
observations that are close to threshold; the estimation of threshold, on the other hand,
might also be affected if there is a spurious peak in the tail (Hu 2013). MacDonald (2011)
showed that uncertainty with threshold choice has a strong localized effect on the estimates
close to the threshold, so the threshold tends to be chosen around the peak in the tail (Hu
2013). Moreover, as Behrens indicated in his paper, the choice of distribution under
threshold can be tricky. If the bulk model distribution is not a good fit, the tail estimation
can be affected as well since the tail model uses distribution information from the overall
observations.

Carreau and Bengui (2008) extends Behren's study and propose a hybrid Pareto model by splicing a normal distribution with generalized Pareto distribution (GPD) and set a continuity constraint on the density and on its first derivative at the threshold (Hu 2013). The fundamental idea of this model is to make hybrid Pareto model not only continuous but also smooth.

## 3.1.c Non-parametric form of Mixture Model

MacDonald (2013) uses kernel density estimators to estimate the non-extreme value distribution and GPD to estimate the tail distribution. Nonparametric kernel density estimators using symmetric kernels work well with populations with unbounded support, or at least a proper tail before the lower boundary (MacDonald 2013). A boundary-corrected kernel density estimator is also used as an alternative for the population with bounded support. This kernel density estimator assumes a particular kernel such as the normal density, which is centered at each data point, and is parameterized by a single bandwidth. MacDonald (2011) also uses the standard cross-validation likelihood to choose the bandwidth, combing with the likelihood for the peaks over threshold tail model, to give a full likelihood for all of the observations. The term tail fraction refers to the proportion of the distribution above the threshold. The notation $\phi_u$ will be used for this quantity when u is the threshold.

The distribution function used by MacDonald takes the form,

$$F(x|\,X,\lambda,u,\sigma_u,\xi,\phi(u)\,) = \begin{cases} (1-\phi(u))\,(\frac{H(x|X,\lambda)}{H(u|X,\lambda)}) & if\, x \leq u \\ (1-\phi(u)) + \phi(u) \times G(x|u,\sigma_u,\xi) & if\, x > u. \end{cases} \quad (2)$$

H(.|X,λ) is the distribution function of the kernel density estimator and λ is the bandwith (Hu 2013). By using nonparametric approaches to estimate the tail risk, the tail fit becomes more robust to the bulk fit, as shown by MacDonald (2013). He uses a sensitivity curve to show the robustness of tail fit to that of the bulk and vice versa. However, while this method makes the estimation more robust, the computational complexity also increases. If we use cross-validation methods for a large sample observation, the overall computing time can be very long.

More importantly, as suggested by Hu (2013), the cross-validation likelihood based kernel bandwidth is biased high in the face of heavy tail behavior. Bowman (1984) suggests that the problem of poor performance is due to the inconsistent likelihood bandwidth parameter.

Despite the benefits of this nonparametric model, it does have some disadvantages. First of all, using cross-validation and kernel estimation increases the computation burden significantly for the large samples. If we use this model to estimate financial data across decades, the underlying computation also increases the model's complexity and computation hours slow down the process of decision making. More importantly, a fundamental assumption of this model is that the distribution has a lower tail that decays to 0 at the boundary. If a distribution that its lower tail does not decay to zero, this model cannot be applied to estimate the tail risk.

### 3.1.d Semi-parametric form of Mixture Model

Nascimento et al 2011 introduces a semi-parametric form of mixture model. He includes a weighted mixture of gamma densities for the bulk distribution and uses GPD as the tail distribution (Hu 2013). By using AIC and BIC, he determines the number of gamma components for the bulk distribution. Compared to the Behrens' approach, the semi-parametric model is relatively more flexible and it overcomes the identifiability problem because of a number of gamma densities in the model. This model's distribution function is taken to be of the form

$$F(x|u, \sigma u, \xi, \eta) = \begin{cases} H(x|\eta) & if x \leq u \\ [1 - H(x|\eta)]G(x|u, \sigma u, \xi) & if x > u. \end{cases} \quad (3)$$

Similar to Behrens's approach model, $H(x|\eta)$ is the conditional bulk distributions of X under the threshold. In this particular case, Nascimento assigns H as gamma function and $G(x|u, \sigma u, \xi)$ represents the GPD density function. $\theta$ represents a shape parameter $\alpha$ and mean paramter $\eta$, of which are any positive value. In Nascimento's paper, he concludes that this weighted mixture of gamma densities can eliminate any constraint on the possible shape of the bulk model and eliminate the unimodal constraint. However, it also uses up less degrees of freedom compared to the non-parametric density estimator (Hu 2013). However, like the model from Behrens (2004), this model would suffer serious drawbacks if the bulk model is mis specified.

## 3.2 Extreme Value Mixture Methods Overview

### 3.2.a Two-step Preprocessing Method

McNeil and Frey (2000) constructed a classic but straightforward model to address the dependence issue. The basic idea is to use GARCH models on the raw data set to extract residuals. They make minimal assumptions about the underlying innovation distribution.

Then they apply an extreme value model on the extracted residuals to estimate tail distributions. The following two-step description is cited from McNeil and Frey's paper.

1. Fit a GARCH model to the return data making no assumption about $F_z(z)$ where z represents the innovations. Estimate $\mu(t+1)$ and $\sigma(t+1)$ using the fitted model and calculate the GARCH model residuals.
2. Consider the residuals to be a realization of a white noise process and use extreme value theory (EVT) to model the tail of $Fz(Z)$ and estimate the tail quantile.

The importance of McNeil and Frey's research opens a new way to estimate the tail risk by analyzing extracted residuals rather than return series itself. With the combination of a GARCH model, this approach effectively solved the independence concern of the classic extreme value theory. This general approach can be extended to the non-parametric model as well. For example, we can use kernel estimation for the bulk model and then using GPD on the extracted residuals. The performance of this method has been widely examined through the implementation of real-life data. However, no study has been conducted to compare this method's performance in for simulated data that have dependences and correlations and combine this method with different extreme value models. In Chapter 5, we will show these methods indeed improves model performance significantly.

### 3.2.b General Extreme Value Model Form

There are two approaches to specifying the tail fraction $\phi_u$ for all extreme value models mentioned above. The first way is to use the parameters from bulk model to define the tail fraction, which is called bulk model-based tail fraction approach (BTF). Another way is to treat $\phi_u$ as a separate parameter to be estimated.

**Bulk Model based tail fraction approach**

Bulk model based tail fraction approach takes the same form of Behrens' approach shown in equation (1). $\phi_u$ represents the tail fraction of the model where $\phi_u$ ranges from 0 to 1. $H(x|\theta)$ represents the parametric distribution of bulk model such as gamma, weibull, normal, lognormal. $G(x|u, \sigma_u, \xi)$ is the GPD parts for the extremal value theorem to estimate the tail fraction.

**Parameterized Model based tail fraction approach**

Parametrized tail fraction approach derives from the MacDonald (2012) study and this approach takes the same form of distribution functions shown in equation (2). This approach uses conditional GPD to estimate the tail distribution estimation. Note that conditional modeling the upper tail requires the proportion of excesses to obtain the unconditional quantities of interest (Hu 2013).

**3.2.c Continuity Constraint Model**

There is no standard framework to connect bulk and tail model. The discontinuity may exist between bulk and tail models at the threshold. Extra continuity constraint models increase the flexibility of the model by enforcing additional mode assumption of models. However, whether such extra continuity constraints will improve the model performance stays unknown.

# Chapter 4 Data and Methodology

## 4.1 Data

To analyze the performance of different models, we will compare their performances in the financial and insurance data. For the financial datasets, we focus on the financial crisis time of 2008. We will use three datasets between 2000 and 2010 to compare the performances of different models and methods. They are the Standard & Poors (SP500), the DAX index, and the price of gold during those periods. SP500 represents the US stock market's performance, which is known for higher volatility and sensitivity to news and the changes in the whole economy. The DAX index is a blue-chip stock market index consisting of 30 major German companies trading on the Frankfort Stock Exchange. Like the SP500, DAX index reflects the stock performances during the financial crisis period and in the European market. The gold price represents a commodity market, which is known for a "safe card" during the financial crisis periods. The underlying reason to apply models in those markets is to find the sensitivity and performances results of different models on various real-life data.

We use a classic Danish Fire Insurance data to analyze potential insurance loss. The Danish Fire Insurance data contains 2167 fire losses in Denmark from January 1980 to December 1990. This data set is already available in the *evmix* library in R. The Danish Fire Insurance is rescaled by -1 and a small residual has been added to it so the data starts above 0 (Hu 2018). Danish fire data is extensively studied during literature due to its fat-tailed features.

The ultimate goal to use such a diversity of datasets is to show how the performances, precision, and accuracy for different extreme value mixture models and methods.

## 4.2 Methodology

**4.2.a Overview**

As we have explored key extreme value models and methods that have appeared in the literature, we find that comparative studies for these models and methods are very scarce. Moreover, there has not been a general framework to optimize model selections based on different types of distributions. Parametric, semi-parametric, and non-parametric models all have their advantages and disadvantages. However, a general framework of choosing those models and apply them into real-life is worth investigating and developing. Chapter 5 and Chapter 6 dig into the applications and performances of these models in depth by cross validating, testing, and measuring their performances on different datasets.

**4.2.b R Package Application**

A comprehensive and useful R package "evmix" that has been developed by Hu (2013) includes most of leading implementations of current extreme value models, which allows for an objective threshold estimation and uncertainty quantification. This package also develops a plethora of boundary corrected kernel density estimation methods to reduce the bias of standard kernel density estimation when a range of support is bounded (Hu 2017). By using this package with additional support from other R packages, we can simulate, test, and measure the performances of different models.

# Chapter 5 Simulation

## 5.1 Simulation Overview

### 5.1.a Methodology

This chapter conducts a detailed comparative study for different extreme value models and methods and examines their performances for tail estimations. We simulate two types of data to compare different models and methods' performances on these two types of data, respectively. The first type of data is constructed by independent and identically distributed random variables, representing ideal datasets that are based on the independence assumption of the extreme value theory. The second type of data is built by identically distributed random variables with correlation and dependency, which is closer to the real-life financial and insurance datasets with volatility clustering.

In general, the overall goodness-of-fit of models is not of interest but rather their accuracy and consistency of tail quantile estimations. Nevertheless, an inappropriate model to fit bulk part of distribution might lead to misleading estimations. In this simulation study, we extend Hu (2013)'s work to investigate the fitness of tail distributions of these models and methods. We simulate different populations with symmetric and asymmetric behaviors that have 3 types of lower and upper tail features. Those features are generally determined by the shape $\xi$ parameters. To give a quick review, the shape parameter is important to determine the behavior of fitting generalized Pareto distribution to the tail distribution.

- Type I: $\xi = 0$: exponential tail
- Type II: $\xi > 0$: heavy tail
- Type III: $\xi < 0$: short tail with the finite upper end point

In our simulation study, we replicate 500 times sampling from each distribution with a sample size of 1000. We estimate 0.1%, 1%, 5%, 10%, 90%, 95%, 99%, and 99.9% tail quantiles and compare estimates with true quantiles to evaluate each model's performance. Since most models are only effective for estimating a single tail (upper tail or lower tail), the poor performance of the other tail's estimation is to be expected. Though evmix R package allows users to choose to focus on lower or upper tails for some models by flipping the tail distribution's estimation, some models are less effective for the other side of the tail. For example, normGPD model that fits the bulk model with normal distribution and the upper tail with Generalized Pareto distribution is only effective to estimate the upper tail of the distribution. For models like normGPD, we will include two tail's estimations, but we should be aware of poor performances for the other tail estimations. To quantify performance estimation, we use the root of mean squared error over all 500 replicates of each model. Let $q_z$ be the true quantile of a distribution and $\hat{q}_z$ be the estimated quantile.

$$RMSE = \sqrt{\sum_{i=1}^{N} \frac{(q_z - \hat{q}_z)^2}{N}}$$

A comprehensive comparative study for extremal models is very scarce in the literature. The most recent work was conducted by Hu (2013) as he compared the performance of different models and also analyzed the model flexibilities with or without a continuous threshold constraint. However, he does not include a lognormal model or a semi-parametric model proposed by Cabras and Castellanos (2011). A classical two-step method for preprocessing datasets with dependence proposed by McNeil and Frey has not been applied to all models that we mentioned in this thesis. The next three sections describe and illustrate simulation results for both independent and dependent datasets.

### 5.1.b Model Abbreviations

We mainly examine 16 models with continuous constraint and two-step methods for our simulation results.

**Single Tail Models**

- normGPD : the normal GPD mixture model
- normGPDcon: the continuous constraint normal GPD mixture model
- normGPD GARCH: the norm GPD model with GARCH method
- hybrid GPD: hybrid Pareto model
- hybrid GPDcon: the continuous constraint hybrid Pareto model
- WeibullGPD: the weibull GPD model
- WeibullGPD GARCH: the weibull GPD model with GARCH method
- GammaGPD: the gamma GPD model
- GammaGPD GARCH: the gamma GPD model with GARCH method
- LognormalGPD: the lognormal GPD model

- o  LognormalGPD GARCH: the lognormal GPD model with GARCH method
- o  KernelGPD: the kernel density estimator GPD model
- o  KernelGPD GARCH: the kernel density estimator GPD model with GARCH method

**Two Tails Models**

- o  GNG: a model with lower tail and upper tail using GPD. The bulk model is fitted with normal distribution. This model is proposed by Zhao (2010).
- o  GNG Con: the continuous constraint GNG model
- o  GNG GARCH: GNG model with GARCH methods.

We use these abbreviations for analyzing simulation results in the next sections.

## 5.2 Simulations for Independent Random Variables
## 5.2.a Generating Independent and Identically Distributed Random Variables

It is straightforward to use random number generators directly in R to simulate independent and identically distributed random numbers. In this section, we simulated 6 populations that represent 3 types of shape parameter $\xi$, namely are

- Norm(0,4): normal distribution with mean = 0 and standardized deviation = 4
- Student(5,0): Student distribution with mean = 0 and degree of free =5
- Student(5,5): Student distribution with mean = 5 and degree of free =5
- Gumbel(5): gumbel distribution with scale = 5

- Weibull (5,2): Weibull distribution with shape =2 and scale = 5

- Reverse Weibull (5,2): Reverse Weibull distribution with shape =2 and scale = 5

These densities of these simulated population are shown in the Figure 5.1 below.

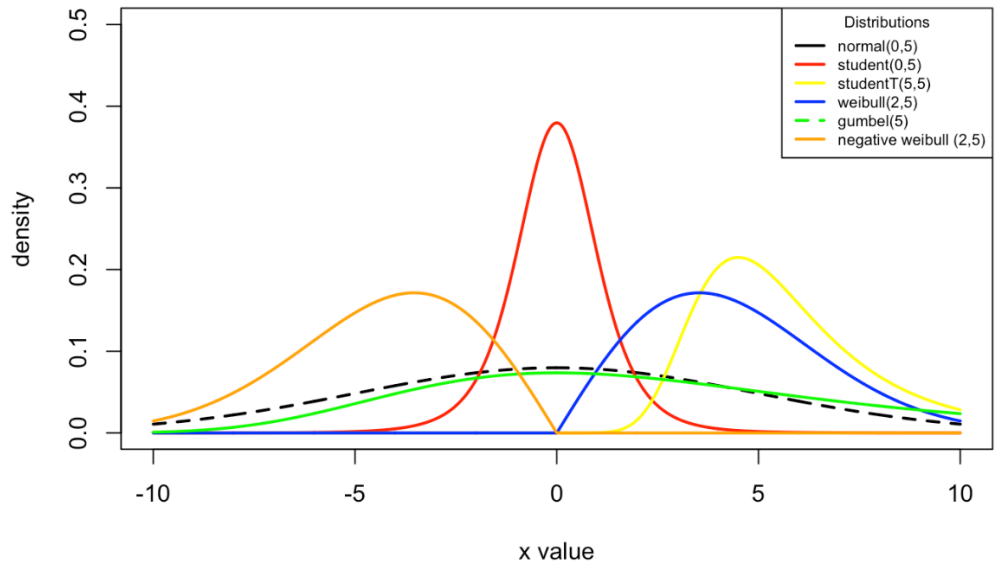

Figure 5.1: Simulated Population for Independent Random Variables

## 5.2.b Results

Table 5.1 shows the simulation results, and it shows some interesting findings. Two-tail models generally have better performance than single tail models. One potential interpretation of this result could be that two-tail models reduce more non-extremal noises from the bulk models and provides a more accurate estimation for the tail distribution. The kernel density estimators based on mixture models do not have an absolute superior or close to the best performers. While it does have a consistently smaller RMSE for symmetric form distribution like Student and the normal distributions, it does not show a significant improvement for estimating extremal quantiles in a heavy-tailed distribution like Gumbel distributions, especially for the estimation of the lower tail. Instead, the other two-tail models like GNG sometimes give similar or even better result at 99.9% or 0.1% quantile. The kernel density estimator with GPD has increased much computing burden. Our findings indicate that for the asymmetric form of distributions, two tail model, GNG, may give a less computing burden and better or close estimation results. Moreover, the continuous constraint method does not show the significant improvement of models. We compare models with and without continuous constraints, and the results are mixed. In the case of single-tail models, the continuous constraints improve the model performance at the lower tail but not at the upper tail. In the case of two-tail models, continuous constraints sometimes may even increase the error rate significantly. This is probably due to the mis-specified bulk model at the threshold. According to the result, if the bulk model is mis-specified like fitting hybrid GPD on the normal distribution, the overall model goodness of fit can be dramatically affected.

|  |  | 0.10% | 1% | 5% | 10% | 90% | 95% | 99% | 99.90% |
|---|---|---|---|---|---|---|---|---|---|
| **Norm(0,4)** | NormGPD | 1.3423752 | 0.489187 | 0.2250154 | 0.1812739 | 0.5247925 | 0.5353045 | 0.5827285 | 0.7981456 |
|  | NormGPDcon | 1.3118155 | 0.4899324 | 0.2228431 | 0.1659903 | 0.5089246 | 0.7194507 | 1.3023527 | 2.2430265 |
|  | Hyprid GPD | 2.3982366 | 1.6041211 | 0.8208743 | 0.629078 | 9.7062 | 14.3910339 | 26.2200162 | 44.5643594 |
|  | Hyprid GPDcon | 2.165224 | 2.271703 | 2.309383 | 2.433811 | 41.211513 | 56.929026 | 94.209597 | 149.194719 |
|  | Kernel GPD | 1.10757625 | **0.35387717** | **0.07392564** | **0.09083726** | **0.01776187** | **0.04133451** | **0.21355787** | **0.59768925** |
|  | GNG | **0.8281677** | 0.3651644 | 0.1876114 | 0.1402858 | 0.1370157 | 0.1811931 | 0.3675285 | 0.8662227 |
|  | GNGcon | 0.9155208 | 0.4244515 | 0.203173 | 0.1510108 | 0.1516992 | 0.2123342 | 0.4402921 | 1.1451695 |
| **Student(5,0)** | NormGPD | 2.1134279 | 0.8524363 | 0.5362356 | 0.4957046 | 0.4726901 | 0.5150449 | 0.6252049 | 0.9816214 |
|  | NormGPDcon | 3.242113 | 3.255869 | 1.659513 | 3.402976 | 6.986709 | 8.935296 | 1.325506 | 1.906643 |
|  | Hyprid GPD | 2.5697493 | 0.8394011 | 0.1572277 | 0.117453 | 2.2388515 | 3.2880947 | 5.6470445 | 8.8652894 |
|  | Hyprid GPDcon | 2.535215 | 0.9101815 | 0.255606 | 0.1116784 | 7.2942735 | 10.2095806 | 16.9232472 | 26.3157007 |
|  | Kernel GPD | **0.19209615** | **0.11828499** | **0.06457724** | **0.05400704** | 0.05130228 | **0.04703773** | 0.67550205 | **0.25787797** |
|  | GNG | 0.69259842 | 0.18592173 | 0.07528238 | 0.05153374 | **0.04880309** | 0.07598302 | **0.1880223** | 0.6950546 |
|  | GNGcon | 0.78844354 | 0.18906772 | 0.05719765 | 0.03600728 | 0.03465938 | 0.06026952 | 0.18616794 | 0.71596158 |
| **Student(5,5)** | NormGPD | 2.0791315 | 0.8241089 | 0.5436467 | 0.4891507 | 0.4385421 | 0.48406 | 0.601526 | 1.0334844 |
|  | NormGPDcon | 1.0762759 | 0.5532443 | 0.2533868 | 0.1627526 | 2.5157184 | 3.6648439 | 7.0032928 | 13.8828561 |
|  | Hyprid GPD | 0.42177774 | 0.16110735 | 0.09794318 | 0.10842433 | 0.96435026 | 1.25183802 | 2.04620616 | 5.3388895 |
|  | Hyprid GPDcon | 0.3865006 | 0.1347113 | 0.1615969 | 0.1925737 | 6.4442677 | 8.7750542 | 13.586635 | 19.5721241 |
|  | Kernel GPD | **0.38265341** | **0.20319909** | **0.14899776** | **0.09583726** | **0.11460668** | **0.09310353** | **0.10736992** | 3.7573493 |
|  | GNG | 0.5706755 | 0.2577925 | 0.1629102 | 0.1368516 | 0.419999 | 0.6651216 | 1.0498539 | **3.2713264** |
|  | GNGcon | 0.3156828 | 0.27842 | 0.212604 | 0.1487842 | 0.3221003 | 0.5931972 | 2.6843076 | 114.958283 |
| **Gumbel(5)** | NormGPD | 0.5685637 | 0.27752964 | 0.12797528 | 0.09959863 | 0.95946017 | 1.15671426 | 1.32220325 | 1.17164016 |
|  | NormGPDcon | 0.70220817 | 0.34541671 | 0.13847018 | 0.06827982 | 0.83813296 | 1.19945771 | 2.12418008 | 3.33554124 |
|  | Hyprid GPD | 0.1548337 | 0.06480339 | 0.06553679 | 0.0578053 | 0.71008433 | 1.12692693 | 2.16873559 | 4.08776461 |
|  | Hyprid GPDcon | 0.17107383 | 0.08910167 | 0.12260061 | 0.12944423 | 4.3236645 | 6.07909229 | 10.2515642 | 16.5029076 |
|  | Kernel GPD | 0.2594442 | 0.12599058 | **0.04894766** | **0.06229197** | **0.01061722** | **0.04485702** | **0.14475577** | **0.40341677** |
|  | GNG | **0.11725889** | **0.07529315** | 0.06330398 | 0.07159941 | 0.15976807 | 0.2294228 | 0.25248335 | 0.49584368 |
|  | GNGcon | 0.1327666 | 0.12840659 | 0.07834592 | 0.04454706 | 0.11000866 | 0.16188647 | 0.32402868 | 1.2863929 |
| **Weibull (5,2)** | NormGPD | 0.12169088 | 0.040081 | 0.03059563 | 0.02799091 | 0.08670148 | 0.07716131 | 0.06465476 | 0.07206183 |
|  | NormGPDcon | 0.11639859 | 0.04338706 | 0.02964748 | 0.02663123 | 0.13657646 | 0.19164473 | 0.32078788 | 0.55008833 |
|  | Hyprid GPD | 0.20776527 | 0.17189494 | 0.07776192 | 0.03068935 | 0.70262993 | 1.10526031 | 2.11778309 | 3.68016124 |
|  | Hyprid GPDcon | 0.18983398 | 0.16714931 | 0.11153143 | 0.08913056 | 2.86825006 | 4.06547195 | 6.91959337 | 11.1193188 |
|  | Kernel GPD | **0.054631407** | **0.023539489** | **0.01149236** | **0.00265309** | **0.01423772** | **0.00781373** | **0.00117815** | **0.02611138** |
|  | GNG | 0.05526414 | 0.03094263 | 0.02419215 | 0.02302487 | 0.02021487 | 0.0193784 | 0.02714645 | 0.04749648 |
|  | GNGcon | 0.06044601 | 0.03738347 | 0.02350952 | 0.01859282 | 0.01568291 | 0.02758566 | 0.05155838 | 0.06498053 |
| **Reverse Weibull(5,2)** | NormGPD | 0.7507936 | 0.4616505 | 0.3589316 | 0.3738832 | 0.6608278 | 0.4652562 | 0.2095265 | 0.0924659 |
|  | NormGPDcon | 0.7478359 | 0.4626243 | 0.3070374 | 0.3025644 | 1.7520417 | 2.1376453 | 2.7150828 | 3.1165986 |
|  | Hyprid GPD | 2.208513 | 1.669794 | 1.116582 | 0.885948 | 7.391823 | 10.644671 | 18.716218 | 30.832605 |
|  | Hyprid GPDcon | 1.4877301 | 0.7721622 | 0.3881498 | 0.3341507 | 31.4208067 | 43.5733503 | 72.3043809 | 113.993339 |
|  | Kernel GPD | 0.50378145 | 0.31086055 | **0.1781492** | 0.07689466 | 0.10166737 | 0.1050239 | 0.06907777 | 0.01803445 |
|  | GNG | **0.4686422** | **0.26898819** | 0.25967073 | 0.24641575 | 0.17900577 | 0.12962645 | 0.07398964 | 0.0609231 |
|  | GNGcon | 6.91622951 | 0.64700967 | 0.32764808 | 0.18984332 | 0.12898007 | 0.19491235 | 0.15850942 | 0.07863296 |

Table 5.1: Simulation results for independent random variables.

Here we sample from different distributions, and the resulting RMSE is computed from the extreme quantile estimation methods.

## 5.3 Simulations for Dependent Random Variables

This section aims to compare the model performances with and without GARCH preprocessing methods. We simulate dependent and identically distributed random variables for different populations and then fit different extreme value models on these simulated datasets.

## 5.3.a Generating Dependent and Identically Distributed Random Variables

In R it is easy to generate independent and identically distributed variables. However, it is challenging to generate correlated random variables from identical distributions. We use the copula method to generate dependent random variables from multivariate distributions. Copulas are popular tools for modeling and simulating data. In the real world, due to the limited data availabilities or missing data, researchers sometimes fit copulas to the returns of financial data to capture underlying correlations and to create datasets more accessible to conduct further studies.

To explain how copulas generate dependent random variables from an identically distributed distribution with a given marginal distribution, we first sample $X_1, X_2, X_3 \ldots X_{1000}$ from a multivariate normal distribution with each $X_i \sim N(0,1)$ but having dependence (using a randomly generated correlation matrix). Then we take $X_i = F^{-1}(\Phi(X_i))$, i=1,2,3…1000, where F is the desired marginal cumulative density function. Note that the correlation matrix has to be semipositive to generate dependent random variables from a multivariate normal distribution. We also use random variables uniform distribution ranging from -0.85 to 0.85 to simulate potential correlations in the real life.

Figure 5.2 gives a panel illustration for three dependent random variables from dependent Gamma (2,1) multivariate distribution.

In this section, we simulated 500 dependent random variables with the sample size of 1000. We compare the RMSE by using the model with and without GARCH preprocessing. The performance of GARCH preprocessing has been studied extensively by backtesting the tail estimations of real-life datasets. However, our study is the first to compare extreme mixture models' performance on large simulated correlated datasets where the underlying true quantiles are known.

We are also curious about comparing the performance of using GARCH preprocessing and that of kernel density estimator GPD model. Kernel density estimator GPD model still has superior performance on the common extremal quantiles like 90% and 95% quantiles, as shown in the previous section. If the two-step method could give a close or even better tail estimations, we can relieve much computing burdens by using the kernel density estimator GPD model.

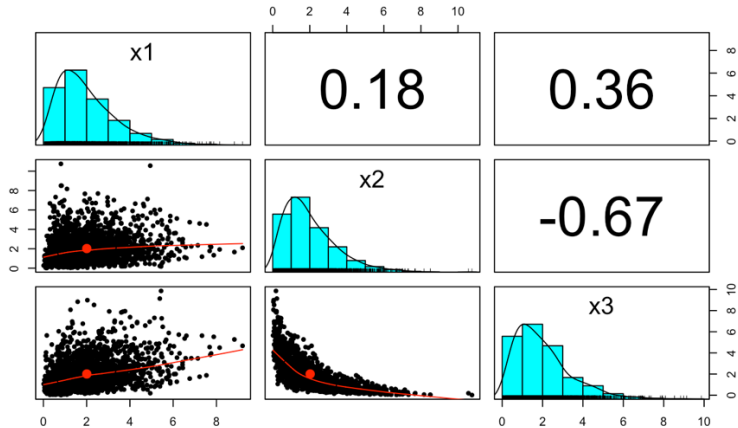

Figure 5.2: Copula Examples with Gamma (shape=2, scale=1)

In this section, we simulated 4 populations.

- Norm (0,4): correlated normal distributions with mean = 0 and standardized deviation = 4

- Student (5,0): correlated Student distribution with mean = 0 and degree of free =5

- Weibull (5,2): correlated Weibull distribution with shape =2 and scale = 5

- Gamma (5,1): correlated Gamma distribution with shape =5 and scale = 1

Assume these simulated datasets are financial time series data. As shown on Figure 5.3, these simulated datasets indeed show the volatility clustering phenomenon.

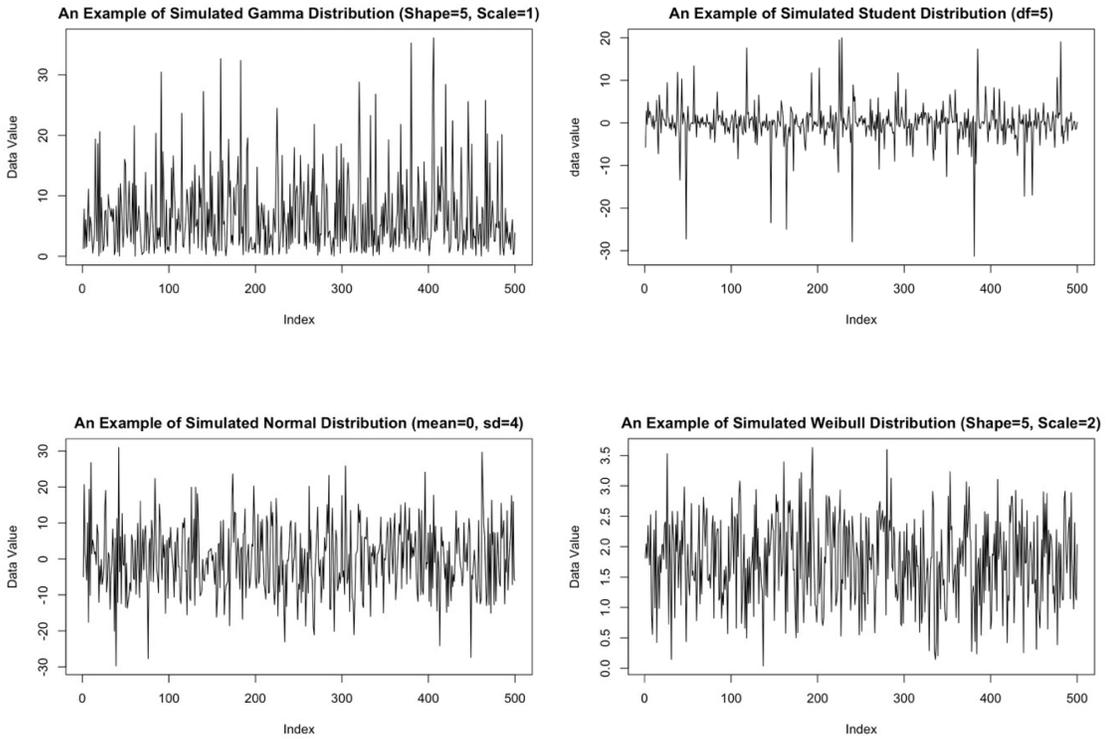

Figure 5.3: Simulated Populations for Dependent Random Variables

**5.3.b Results**

The simulation results are shown in Table 5.2. Preprocessing dependent datasets with GARCH models significantly improves the models' performances. The error rates are reduced by preprocessing datasets with GARCH. Since original datasets are highly correlated, the key assumption of the extreme value theory is violated. The residuals from GARCH models solve the violation of the dependence. Estimating tail distributions based on extracted residuals from GARCH, therefore, improves the models' performances. The results for comparing kernel density estimator GPD and other models with GARCH are mixed. Kernel density estimator GPD alone does not show superior performances compared to the GNG or other models with GARCH. That is to say, using GARCH method to preprocess data can get even better results compared to apply kernel density estimator GPD directly. However, kernel density estimator GPD with GARCH indeed has better performance for the symmetric distributions like normal distribution. For asymmetric distributions like gamma and Weibull distribution, the models that describe the bulk model distributions better give better results. It again shows the importance to choose appropriate bulk models.

In summary, GARCH methods significantly improve model performances, and even give better results than kernel GPD model. For estimating common tail quantiles estimation, preprocessing the datasets by GARCH and choosing appropriate bulk models can give a significant improvement for tail distribution estimations.

| | RMSE | 0.10% | 1% | 5% | 10% | 90% | 95% | 99% | 99.90% |
|---|---|---|---|---|---|---|---|---|---|
| | NormGPD | 4.738458 | 2.796808 | 1.189241 | 0.524543 | 1.473969 | 1.992446 | 3.444503 | 14.311206 |
| | NormGPD GARCH | 1.244376 | 0.762046 | 0.355628 | 0.163039 | 0.299117 | 0.525017 | 0.595262 | 0.9948535 |
| | Hyprid GPD | 4.747586 | 2.712758 | 0.977132 | 0.249879 | 7.821454 | 10.43749 | 16.06365 | 21.5856164 |
| | Hyprid GPD GARCH | 0.437617 | 0.231807 | 0.067006 | 0.010497 | 0.671709 | 0.918548 | 1.364572 | 1.2007295 |
| | GNG | 0.001783 | 0.009878 | 0.08823 | 0.282182 | 2.469359 | 4.674763 | 4.325734 | 2.081007167 |
| | GNG GARCH | 0.001158 | **0.004264** | 0.018482 | 0.042808 | 0.322558 | 0.598189 | 0.550125 | **0.27757049** |
| | WeibullGPD | 0.001652 | 0.014969 | 0.038451 | 0.042961 | 0.829412 | 1.03153 | 1.280877 | 3.483392118 |
| | WeibullGPD GARCH | 0.00081 | 0.004883 | 0.00891 | 0.010492 | 0.077483 | 0.11133 | **0.161871** | 0.341836293 |
| | kernelGPD | 3.03195 | 1.756541 | 0.722093 | 0.291133 | 0.415015 | 0.957315 | **1.398893** | 2.0471016 |
| | KernelGPD GARCH | 0.455145 | 0.255086 | 0.105961 | 0.055366 | 0.036106 | 0.143518 | **0.182274** | 0.34267726 |
| | LognormalGPD | 0.003032 | 0.011194 | 0.043344 | 0.082466 | 0.846022 | 0.98723 | 1.683693 | 11.62575501 |
| | LognormalGPD GARCH | 0.004056 | 0.008585 | 0.007586 | 0.011151 | 0.104434 | 0.121099 | 0.216177 | 1.269372483 |
| | GammaGPD | 0.001335 | 0.008996 | 0.023847 | 0.038659 | 0.602916 | 0.922087 | 1.269229 | 3.219640748 |
| Gamma(5,1) | GammaGPD GARCH | **0.000808** | 0.004566 | **0.00678** | 0.009434 | 0.066428 | 0.099559 | 0.165325 | 0.371487356 |
| | NormGPD | 762.7326 | 188.4381 | 112.4706 | 50.29422 | 8.875941 | 16.63686 | 86.11691 | 1335.927437 |
| | NormGPD GARCH | 14.78823 | 8.706457 | 3.859469 | 2.760517 | 0.645573 | 0.152846 | 0.901803 | **5.7307713** |
| | Hyprid GPD | 899.6225 | 133.7884 | 114.8942 | 84.6123 | 443.9525 | 601.1457 | 941.1156 | 1347.7499 |
| | Hyprid GPD GARCH | 6.751608 | 3.772725 | 2.173037 | 2.126793 | 16.60669 | 23.83456 | 41.0851 | 64.676626 |
| | GNG | 8.84E+03 | 1.57E+02 | 8.11E+00 | 3.99E+00 | 2.07E+00 | 2.41E+00 | 1.10E+03 | 7.95E+06 |
| | GNG GARCH | 6.29E-01 | 9.58E-01 | 1.92E+00 | 2.58E+00 | 1.89E+00 | 1.05E+00 | 3.49E+00 | 5.26E+01 |
| | WeibullGPD | 1.16E+03 | 8.61E+01 | 1.31E+01 | 6.25E+00 | 4.38E+00 | 1.19E+01 | 4.79E+01 | 6.01E+02 |
| | WeibullGPD GARCH | 0.2405 | 0.431739 | **0.432126** | 0.743818 | 0.473276 | **0.149234** | 0.966935 | 7.2859992 |
| | kernelGPD | 5.17E+05 | 2.32E+00 | 2.76E+00 | 1.77E+01 | 1.28E+01 | 4.58E-01 | 7.34E+00 | 1.20E+02 |
| | KernelGPD GARCH | 4.382884 | 2.93998 | 1.87855 | 1.228447 | **0.176266** | 0.916736 | 4.425276 | 20.8372509 |
| | LognormalGPD | 1.16E+03 | 8.61E+01 | 1.32E+01 | 6.28E+00 | 8.64E+00 | 2.09E+01 | 7.37E+01 | 1.07E+06 |
| | LognormalGPD GARCH | **0.064629** | 0.350744 | 0.585754 | **0.67517** | 0.675618 | 0.482339 | 0.89083 | 4.61777145 |
| | GammaGPD | 1.16E+03 | 8.61E+01 | 1.32E+01 | 6.27E+00 | 4.99E+00 | 1.09E+01 | 4.40E+01 | 5.80E+02 |
| Student(df=5) | GammaGPD GARCH | 0.176867 | 0.442782 | 0.513017 | 0.888378 | 0.49443 | 0.190786 | 0.886236 | 4.4387693 |
| | NormGPD | 2.94666 | 1.420847 | 0.425665 | 0.582912 | 0.228904 | 0.537716 | 1.440565 | 1.1580276 |
| | NormGPD GARCH | 0.801602 | 0.220308 | 0.090078 | 0.068993 | 0.043215 | 0.062053 | 0.106262 | 0.15208939 |
| | Hyprid GPD | 3.830415 | 2.449128 | 1.266979 | 1.113045 | 19.7533 | 29.08621 | 53.6773 | 91.94144 |
| | Hyprid GPD GARCH | 0.307165 | 0.469736 | 0.217428 | 0.094529 | 1.854504 | 2.909358 | 5.47671 | 9.65995273 |
| | GNG | 2.26E+00 | 1.08E+00 | 5.74E-01 | 3.71E-01 | 2.90E-01 | 5.31E-01 | 1.29E+00 | 9.75E-01 |
| | GNG GARCH | 1.25E-02 | 1.16E-01 | 1.99E-01 | 8.80E-02 | 5.97E-02 | 7.16E-02 | 1.06E-01 | 1.27E-01 |
| | WeibullGPD | 2.87E+01 | 2.23E+01 | 1.68E+01 | 1.40E+01 | 4.35E+00 | 3.34E+00 | 2.38E+00 | 4.25E+00 |
| | WeibullGPD GARCH | **0.234236** | 0.512225 | 0.205298 | 0.090875 | 0.088696 | 0.088044 | 0.094484 | 0.16368938 |
| | kernelGPD | 2.423228 | 1.063729 | 0.526304 | 0.756526 | 0.771549 | 1.248997 | 1.341829 | 0.154076 |
| | KernelGPD GARCH | 0.430901 | **0.053286** | 0.085022 | 0.047182 | 0.056797 | 0.085789 | 0.052523 | **0.06411373** |
| | LognormalGPD | 2.88E+01 | 2.24E+01 | 1.68E+01 | 1.38E+01 | 5.58E+00 | 4.35E+00 | 3.35E+00 | 5.12E+00 |
| | LognormalGPD GARCH | 0.335828 | 0.511345 | 0.149869 | 0.182692 | 0.220402 | 0.214184 | 0.326102 | 1.1838 |
| | GammaGPD | 2.87E+01 | 2.23E+01 | 1.68E+01 | 1.39E+01 | 4.69E+00 | 3.57E+00 | 2.51E+00 | 5.61E+00 |
| Norm(0,4) | GammaGPD GARCH | 0.319693 | 0.558287 | 0.17637 | 0.096177 | 0.187813 | 0.162777 | 0.114277 | 0.20722756 |
| | NormGPD | 4.003378 | 2.395203 | 0.992975 | 0.359556 | 0.572262 | 0.875965 | 1.295037 | 3.0115244 |
| | NormGPD GARCH | 0.889755 | 0.522456 | 0.21542 | 0.089909 | 0.200844 | 0.276798 | 0.27593 | 0.2508407 |
| | Hyprid GPD | 5.035425 | 2.951459 | 1.068028 | 0.195361 | 5.879568 | 7.812003 | 13.9687 | 26.00052 |
| | Hyprid GPD GARCH | 0.676009 | 0.378572 | 0.123407 | 0.020677 | 0.260966 | 0.328695 | 0.807696 | 2.2797576 |
| | GNG | 3.43E-04 | 4.21E-03 | 1.90E-02 | 1.11E-01 | 1.78E+00 | 2.79E+00 | 1.99E+00 | 6.93E-01 |
| | GNG GARCH | 7.39E-04 | **3.47E-03** | **9.27E-03** | 2.79E-02 | 3.69E-01 | 5.33E-01 | 3.62E-01 | 1.09E-01 |
| | WeibullGPD | 3.16E-04 | 7.17E-03 | 3.93E-02 | 5.45E-02 | 1.20E+00 | 8.90E-01 | 9.76E-01 | 1.56E+00 |
| | WeibullGPD GARCH | 0.000601 | 0.006479 | 0.01413 | 0.026061 | 0.202635 | 0.157558 | 0.146711 | 0.212831384 |
| | kernelGPD | 3.38E+00 | 2.12E+00 | 9.23E-01 | 3.20E-01 | **2.43E-01** | 4.84E-01 | 7.36E-01 | 8.48E-01 |
| | KernelGPD GARCH | 0.625948 | 0.391394 | 0.17286 | 0.059045 | 0.042551 | **0.072835** | 0.118948 | **0.19769379** |
| | LognormalGPD | 2.15E-03 | 9.24E-03 | 2.14E-02 | 5.35E-02 | 1.60E+00 | 1.27E+00 | 1.33E+00 | 1.65E+00 |
| | LognormalGPD GARCH | 0.003436 | 0.010657 | 0.011253 | **0.013307** | 0.188608 | 0.151409 | 0.164558 | 0.223767785 |
| | GammaGPD | **1.08E-04** | 3.27E-03 | 2.16E-02 | 3.11E-02 | 1.14E+00 | 8.41E-01 | 9.31E-01 | 1.60E+00 |
| Weibull(5,2) | GammaGPD GARCH | 0.000472 | 0.005856 | 0.011901 | 0.02274 | 0.214014 | 0.147901 | 0.140798 | 0.293963034 |

Table 5.2: Results for Dependent Random Variables

# Chapter 6 Application

This chapter extends the results from the last chapter and implements different extreme mixture models on four real-life datasets.

## 6.1 2000-2010 Financial Crisis

The 2008 financial crisis was the worst economic disaster since the Great Depression in 1929, which led the Great Recession in the U.S economy. The first warning signs of problems in the economy started in 2006 when housing prices began dropping. Banks had been issuing too many questionable credits for mortgage loans. For example, people were allowed to borrow without accompanying appropriate levels of collaterals, and they were unable to pay back these loans (The Balance Website). This overheating housing market was pushed to the edge by mortgage-backed securities that bundled with loans. Mortgage-backed securities were owned by hedge funds, mutual funds, pension funds, and corporate assets. Since borrowers were unable to pay back, banks realized that they had to absorb all of the associated loss. Internal borrowing chain among banking systems then broke since no bank wanted to take those worthless mortgages as collaterals. The whole system crashed in March 2008 when shareholders sold off their shares of investment bank Bear Stearns since it was holding too many "bad" assets backed by those mortgages. Since then, all correlated banking and companies have suffered extremal losses and mistrust to their shareholders. Stock markets suffered a high volume of the selloff and the volatility.

We picked three datasets between 2000 and 2010 from this period to compare the performance of different models. They are the Standard & Poors (SP500), the DAX index, and the price of gold during those periods. SP500 represents the US stock market's performance, which is known for higher volatilities and sensitivities to news and changes in the macroeconomic environment. DAX index is a blue-chip stock market index consisting of 30 major German companies trading on the Frankfort stock exchange. Like SP500, DAX index reflects the stock performances during the financial crisis period but in the European market. The gold price is included to represent a specific commodity market.

We first take a look at three time series plot for these 3 data set between 2000-2010 years. Between 2008 and 2009, we can see volatility clustering and extremal events, especially around 2008.

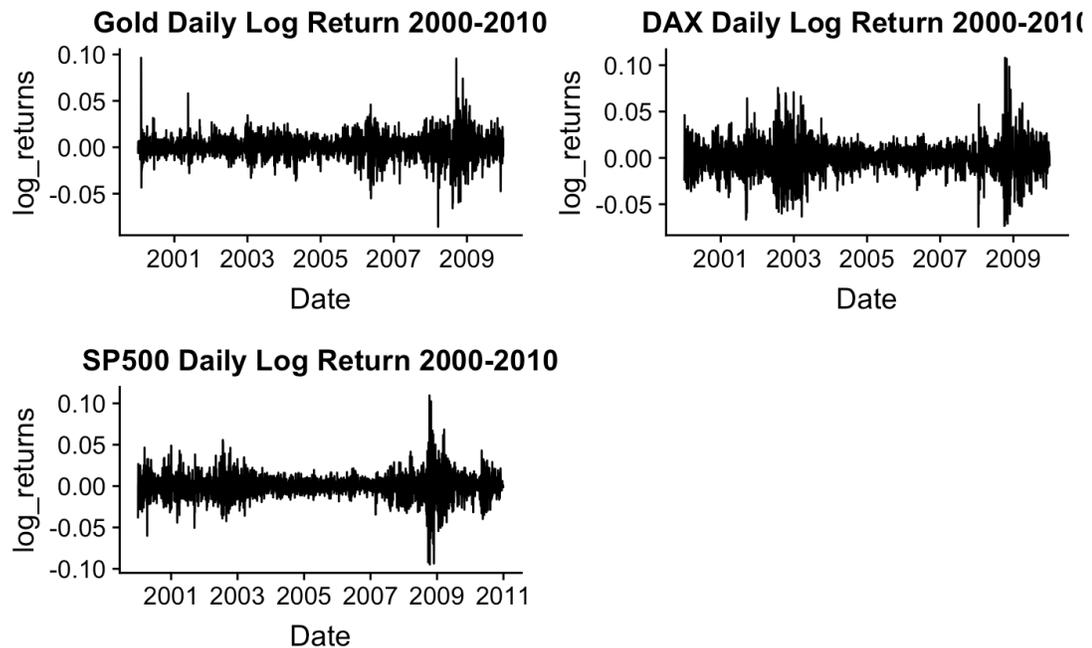

Figure 6.1: Time Series Plot for Gold, DAX, and SP500 from 2000 to 2010

We take log returns of these assets price. Descriptive statistics for these log returns are presented in Table 6.1. We can see that Gold and DAX datasets show the positive skewness indicating frequent small loss and some extreme losses. SP500 shows a relatively strong negative skewness that represents some small gains and extreme gains.

Moreover, the kurtosis coefficients are greater than three times of the normal distribution, indicating that underlying distributions have heavier tails than the normal distribution.

| Statistics | Gold | SP500 | DAX |
|---|---|---|---|
| Mean | 0.000540012 | -5.27E-05 | -4.92E-05 |
| Medium | 0.000434311 | 0.000510416 | 0.000693 |
| Minimum | -0.08571059 | -0.09469514 | -0.074335 |
| Maximum | 0.09641629 | 0.109572 | 0.107975 |
| Standard Dev | 0.01216464 | 0.01377794 | 0.016739 |
| Skewness | 0.03911581 | -0.1133197 | 0.071849 |
| Kurtosis | 6.701688 | 7.520764 | 4.084037 |

Table 6.1: Descriptive Statistics for Gold, DAX, and SP500 from 2000 to 2010

The following Figure 6.2 provides the mean residual life plots for Gold, SP500, and DAX. The thresholds have been chosen as 0.013, 0.014, and 0.018, respectively. We will use these thresholds as the initial values to estimate the models' parameters. Our next step is to use the appropriate extreme value model to estimate the tail distribution.

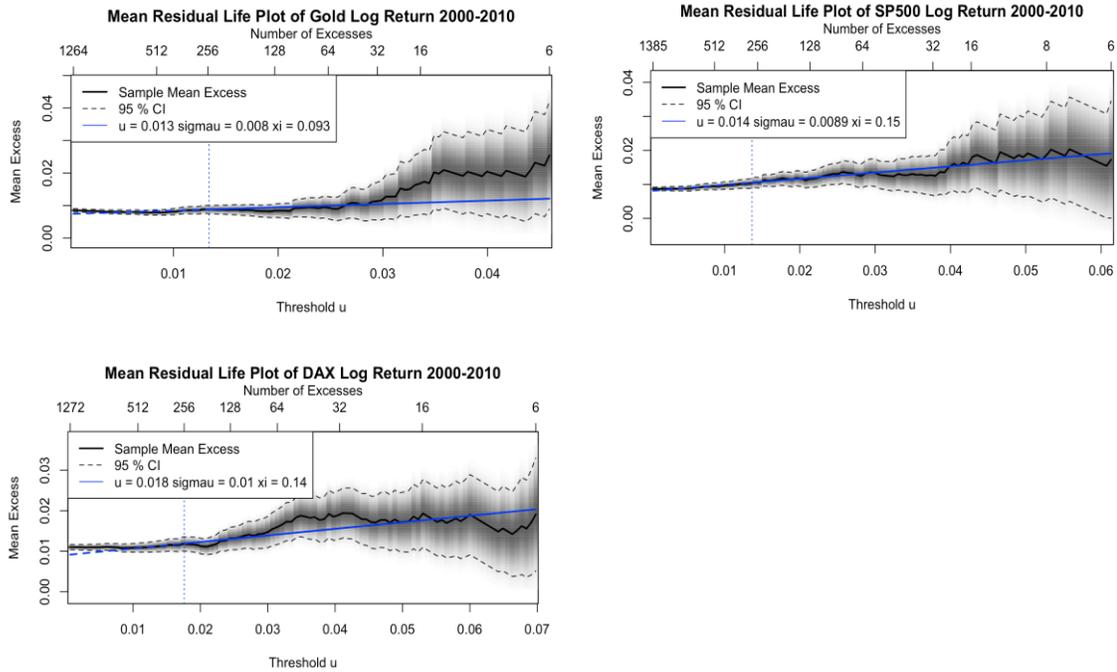

Figure 6.2: Mean Residual Life Plots for Gold, DAX, and SP500 from 2000 to 2010

We fit normGPD, normGPDcon, Kernel GPD, and GNG models on three financial data sets. Figure 6.3 provides the histogram of the data overlaid with the fitted density plot. The bulk-based approach normal GPD and continuous normal GPD fail to capture the mode of the distribution, which underestimates the extremal values of tail distributions. The hybrid GPD model completely mis-specifies the distribution, so we do not include it in our results here. However, as noted in the figure, the GNG model proposed by Zhao (2010) performs the best. To give a quick review, GNG models include the two-tailed normal GPD, that is, both lower and upper tails are modeled by GPD, and the non-extremal middle part is assumed to be normally distributed. While GNG still seems to miss the highest mode of the distribution (the most extremal data), GNG does capture most of the extremal data points.

This result matches with our simulation results that GNG gives a relatively smaller RMSE for heavily-skewed data.

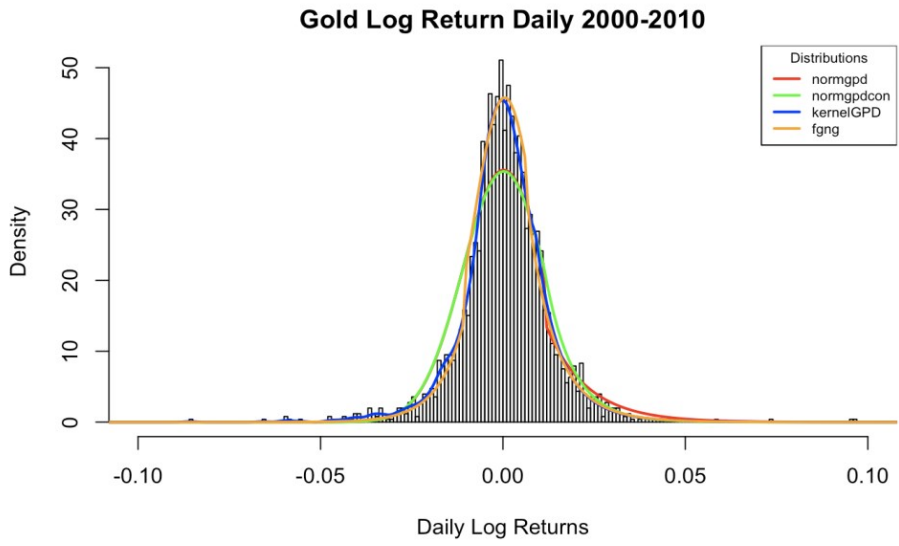

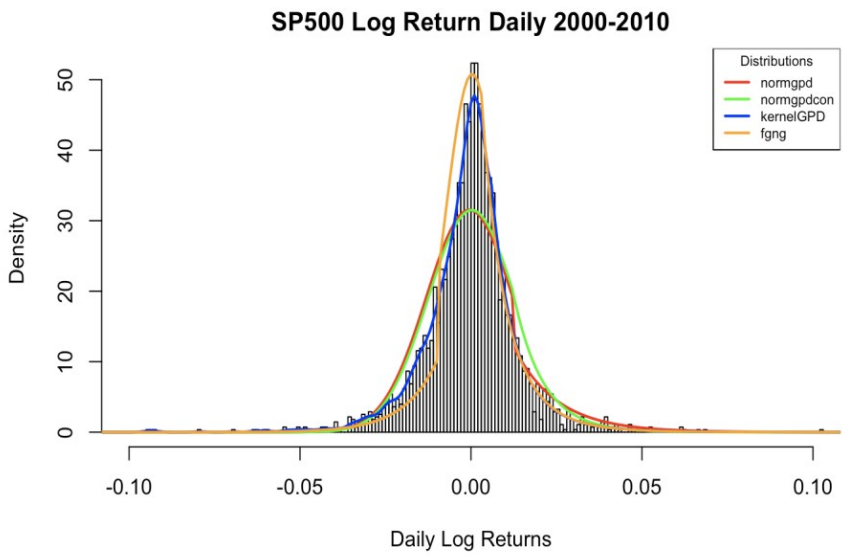

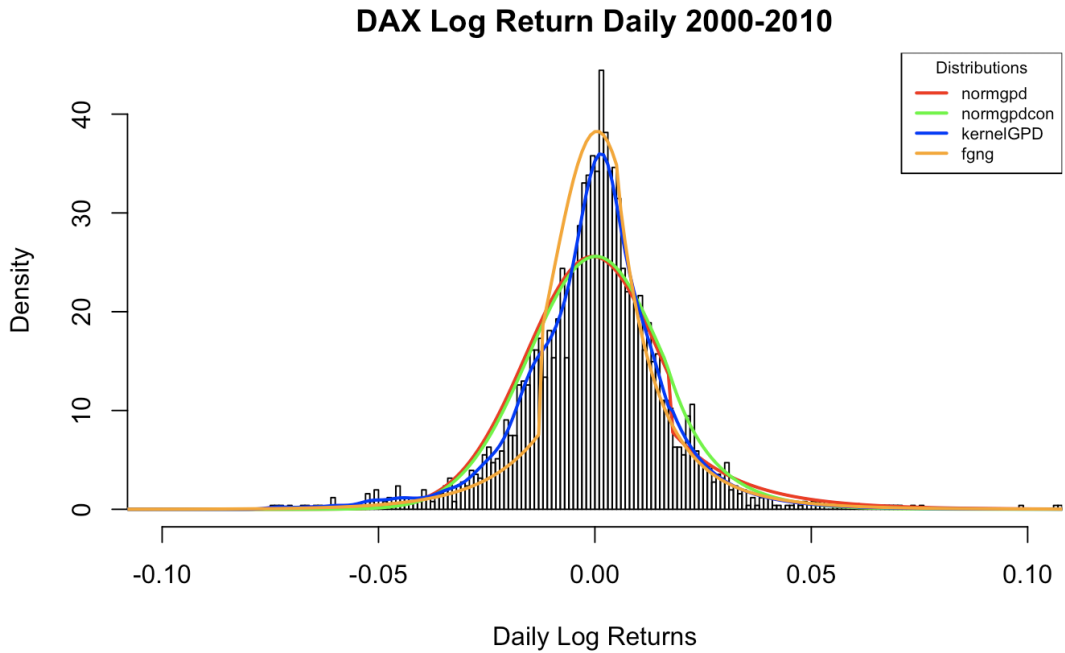

Figure 6.3: Application Results Plots for Gold, DAX, and SP500 from 2000 to 2010

## 6.2 Danish Fire Insurance Data

The Danish Fire Insurance data contains 2167 fire losses in Denmark from January 1980 to December 1990. This data set is already available in the evmix library in R so it is easy to get data access. The Danish Fire Insurance is rescaled by -1 and have a small residual added to it so the data starts above 0 (Hu 2013). Danish fire data is extensively studied during literature due to its fat-tailed features, and it is useful to give a sample for potential insurance claim application. Although Hu (2013) has already conducted a brief analysis on this dataset, he did not implement the lognormal model. In Chapter 5 we have shown that GARCH methods can improve extreme mixture models' performances significantly, especially for heavily-fatted tail distribution. We implement extreme mixture models with and without GARCH to compare their performances.

The Danish Fire Insurance data has extensively studied by McNeil and Embrechts (1999) who used peaks over threshold method to estimate the quantile of the fire losses. Resnick also uses several diagnostics for the independence assumption of the Danish data and concluded that there exists dependence of the data.

After implementing extreme value models directly on the Danish data in log scale, we find none of the models captures all of the tail distribution correctly, as shown in Figure 6.4. Because the tail is heavily left skewed, all models tend to find a further center right from the actual center. We suspect that it is partially due to the highly correlated and dependent on each data point. In order to isolate this potential dependency, we implement a two-step model to preprocess original data by fitting a GARCH (1,1) model on the original dataset.

We fit Garch (1,1) on the original Danish data in log scale and then extracted the residuals. Figure 6.5 shows the result of GARCH fitting results. With the statistically significant coefficients, GARCH (1,1) is an appropriate to preprocess the dataset.

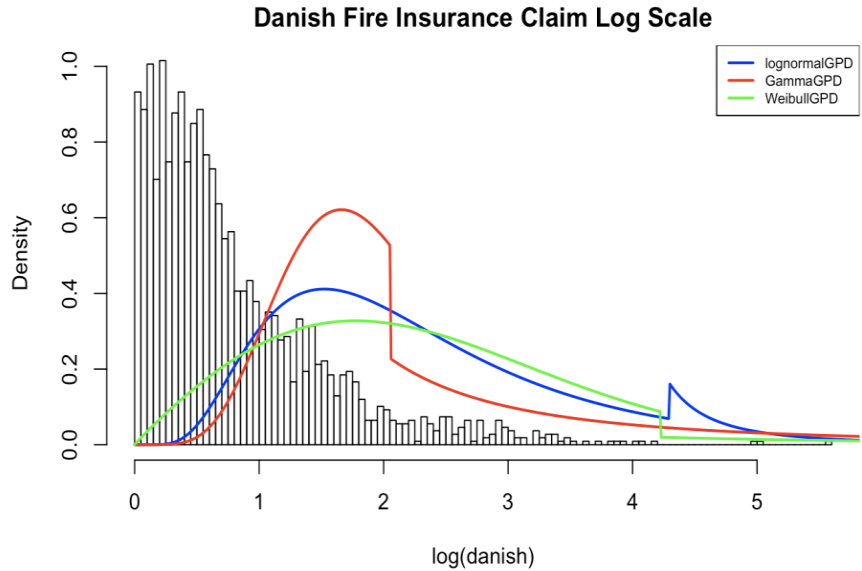

Figure 6.4: Application Results Plots for Danish with GARCH

```
Error Analysis:
        Estimate  Std. Error  t value Pr(>|t|)
mu       0.78648     0.01538   51.130   <2e-16 ***
omega    0.35784     0.16075    2.226   0.0260 *
alpha1   0.02560     0.01337    1.914   0.0556 .
beta1    0.27669     0.31661    0.874   0.3822
---
Signif. codes:  0 '***' 0.001 '**' 0.01 '*' 0.05 '.' 0.1 ' ' 1

Log Likelihood:
 -2349.307    normalized:  -1.084129
```

Figure 6.5: GARCH Results for Danish

After preprocessing the dataset, we apply extreme value mixture models on the residuals to fit the distribution. Figure 6.6 shows that now the estimation for tail distribution improves significantly. The weibullGPD and GammaGPD model successfully capture the extremal portion of the distribution. While both of them seem to overestimate the mode of the heavy tail, model performances have been significantly improved. By showing the better estimation after preprocessing with GARCH (1,1), we have verified that the two-step model method indeed improves the tail estimation of heavily fat tail. The result of the application matches our simulation conclusion.

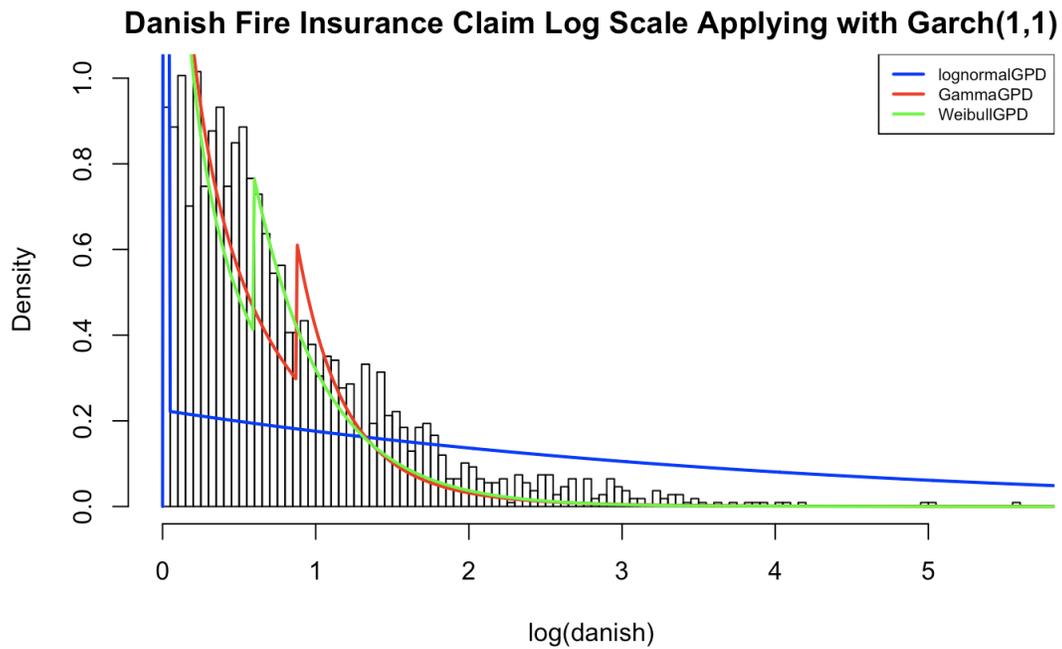

Figure 6.6: Application Results for Danish with GARCH

# Chapter 7 Conclusion

This thesis evaluates the main extreme mixture models and methods and implements them in financial and insurance data sets to compare their performances. The paper also reviews and studies the extreme value theory, time series, volatility clustering, and risk models in details. By extending and using the R package of evmix, this thesis simulates both independent and dependent datasets from identically distributed random variables.

Based on our simulation and application results, we find that two-tailed models, in general, are more effective than the single-tailed models. Two-tailed models apply the GPD model on both upper and lower tails with the assumption that the non-extremal portion is normally distributed.

We find a significant model improvement by implementing the two-step method proposed by McNeil and Frey. Two-step method preprocesses data by fitting a GARCH model and then applies extreme value mixture models on the residuals extracted from the GARCH fit. Extracted residuals are normal and independent distributed, which solves the problem of the dependency that violates the key assumption of extreme value mixture theories.

There are a few other interesting findings. First, the kernel density estimators based mixture models do not have an absolute superior or close to the best performers. While it does have a smaller RMSE for the lower quantile estimation of symmetric form distributions like Student and normal distributions, it does not show a significant advantage for estimating heavy-tailed distributions such as Gumbel and Weibull distributions. Instead, two tail models like GNG have a close or even better result at 99% and 99.9% estimation. The kernel density estimator has increased much computing burden. Secondly, the continuity constraint does not necessarily give a better estimation. We compare the model with and without continuity constraints, and the result is mixed. Continuity constraints sometimes can even increase the error rate. This is probably due to the mis-specified model at the threshold. Last, we find that if the bulk model is mis-specified, the overall model goodness of fit can be dramatically affected.

In summary, this thesis has contributed to providing a systematic comparative study for extreme mixture models and methods. It is also the first work to investigate two-step methods on diverse extreme value mixture models. Further studies can extend to study the effect of outliners for the tail distribution estimation.

# Vita

Yujuan Qiu is from Chengdu, China. She is a graduate student in the combined Bachelor's Master's program in Applied Mathematics and Statistics at Johns Hopkins University. She is particularly interested in financial mathematics and statistics. She likes music, cooking, and painting. She also has 4 cats at home. After graduation, she plans to go to work in the finance industry.